\documentclass[useAMS, usenatbib, fleqn]{mn2e}
\usepackage{graphicx}
\usepackage{natbib} 

\usepackage{txfonts, ulem}
\bibpunct{(}{)}{;}{a}{}{,}

 \title[Super-luminous Supernovae as Quark Novae]
 {Quark Nova Signatures in Super-luminous Supernovae}

   \author[M. Kostka et al.]
   {M.~Kostka,$^1$\thanks{email:mkostka@ucalgary.ca}
     N. Koning,$^1$ R.~Ouyed,$^1$ D.A. Leahy,$^1$ and W. Steffen$^2$\\
    $^1$Department of Physics and Astronomy, University of Calgary, 
2500 University Drive NW, Calgary, Alberta, T2N 1N4 Canada\\             
   $^2$Instituto de Astronom\'{i}a Universidad Nacional Autonoma de M\'{e}xico, Ensenada, B.C., Mexico}

   \date{Received ; accepted }
   
\begin{document}
\label{firstpage} 

\maketitle

  \begin{abstract}
Recent observational surveys have uncovered the existence of super-luminous supernovae (SLSNe).  While several possible explanations have been put forth, a consensus description for SLSNe has yet to be found.  In this work we study the light curves of eight SLSNe in the context of dual-shock quark novae. We find that progenitor stars in the range of 25-35 $M_{\sun}$ provide ample energy to power each light curve.  An examination into the effects of varying the physical properties of a dual-shock quark nova on light curve composition is undertaken.  We conclude that the wide variety of SLSN light curve morphologies can be explained predominantly by variations in the length of time between supernova and quark nova.  Our analysis shows that a singular H$\alpha$ spectral profile found in three SLSNe can be naturally described in the dual-shock quark nova scenario.  Predictions of spectral signatures unique to the dual-shock quark nova are presented.

  \end{abstract}


\section{Introduction}
\label{intro}

 The standard astrophysical explanation for a supernova (SN) is that the radiated power is generated by energy deposited in an expanding ejecta through one of three mechanisms: the SN shock travels through the stellar envelope \citep{grassberg71}, radioactive decay of heavy elements synthesized in the explosion \citep{arnett82} or a collision with hydrogen-rich circumstellar material (CSM) \citep{chev82}. In 2011 astronomers working on the Palomar Transient Factory announced the emergence of a new class of SNe that cannot be explained by any of these means \citep{quimby11}.  As described by \cite{quimby11} this new class of SNe is at least ten times brighter than a typical type Ia SN, displays spectra with little to no hydrogen, emits significant UV flux over a long period of time and has a late stage luminosity evolution that is inconsistent with radioactive decay.  

While this hydrogen-poor class of super-luminous SNe (SLSNe) is recent admission, the phenomenon of SLSNe as a whole has been an open question since the discovery of SN 2006gy \citep{quimby07}.  Large scale supernovae surveys such as the Palomar Transient Factory (PTF) \citep{rau09, law09}, the ROTSE Supernova Verification Project (RSVP, formerly the Texas Supernova Search) \citep{quimby05} and the Catarina Real-Time Transient Survey \citep{drake09a} have uncovered approximately ten other SLSNe, some of which contain hydrogen in their spectra (SN 2006gy \citep{quimby07}, SN 2008fz \citep{drake10}) while others are hydrogen-poor (SN 2005ap \citep{quimby07}, SN 2007bi \citep{galyam09}). In order to describe these powerful explosions our understanding of the evolution of massive stars must change.

\subsection{Proposed SLSN Models}
 \label{proposed}
One method being considered to power the radiated energy of some SLSNe is a scaled up version of a CSM interaction.  A dense, massive ($\sim20$M$_{\sun}$) CSM envelope must enshroud the progenitor star at the time of SN explosion in order for the SN shock to generate enough energy to power the SLSN light curve \citep{smith08,chev11,ginzburg12,kiewe12}.  Building such a CSM envelope requires a mass-loss rate of $\dot{M} > 0.1 M_{\sun} $ yr$^{-1}$ over the final 10-100 years prior to SN explosion \citep{moriya12,ginzburg12}.  Two possible explanations for a mass-loss rate on this order are LBV-like mass ejections \citep{smith08,kiewe12} or common envelope phase of an interacting binary system \citep{chev12}.  In common envelope description, the SN is powered by the inspiral of a compact object on the core of the companion star \citep{chev12}.  As well as requiring a massive CSM envelope,  the conversion efficiency of kinetic energy to radiation must be much higher than that of a typical SN ($\sim 1\%$) in order to power SLSNe light curves \citep{moriya12,ginzburg12}.  Alternatively increasing the kinetic energy of the SN by approximately ten-fold would have a similar effect on the energetics as increasing the conversion efficiency.  The CSM model cannot explain hydrogen-poor SLSN, only type II SLSNe with strong hydrogen lines \citep{kiewe12}.

An alternative description considered for SLSNe is that the radiated energy is converted from the rotational energy of a magnetar \citep{kasen10,woosley10} inside a SN envelope.  For the magnetar model to power the light curve of a SLSN, large $PdV$ losses must be avoided by delaying the conversion of the magnetar's rotational energy into radiation.  The delayed injection of energy into the SN envelope must be isotropically distributed across the inner edge of the SN envelope in order to energize the entire envelope and generate enough radiated energy to power a SLSN.  Whether the magnetar model can power a SLSN with the expected jet-like \citep{Bucciantini09} energy deposition has yet to be studied.  Although the isotropic magnetar model can provide a decent fit to the SLSN light curve, it gives no natural explanation for the observed emission lines.

Pair-instability SNe (PISNe) have as well been proposed as the underlying energy mechanism for SLSNe.  In this scenario an extremely massive star becomes prone to $\gamma = 4/3$ instability, triggering a SN explosion.  \cite{pan12} studied the progenitor stars for PISN and found that the mass range required for a star to end its life as a PISN is $\sim300-1000 M_{\sun}$ and that no star under $1000 M_{\sun}$ in the low red-shift Universe would be susceptible to PISN.  The rise time of the PISN light curve is typically much longer than that observed in the light curves of SLSNe \citep{kawabata09}.  In order to model the light curve of SN 2006gy \cite{woosley07} suggested that pair-instability ejections had occurred twice and that the second ejection caught and crashed into the first ejection.  \cite{woosley07} stated that in order for this collision to occur the kinetic energy of the second ejection must be artificially increased.  PISN was also suggested by \cite{galyam09} as a possible explanation for SN 2007bi because the SN showed no evidence of a CSM interaction.  In the supplemental material associated with the work presented by \cite{galyam09} it was noted that essentially all the available nuclear energy of the progenitor star must be converted to kinetic energy in order for the PISN to explain the light curve of SN 2007bi.  Table \ref{compModelsTable} summarizes each proposed model's explanation for a variety of SLSNe characteristics.

\cite{ouyed02} first suggested that a collision between material ejected through a quark nova (QN) and the preceding SN envelope could rebrighten the SN (see section 5.4 of \cite{ouyed02}).  This theory was first applied in the context of SLSNe by \cite{leahy08}.  In this work we provide the QN as a possible universal explanation for SLSNe.  In turn we discuss what appear to be observational indications that QNe may lay at the heart of SLSNe.  Observations of the SLSNe studied in this work are introduced in section \ref{obs}.  Section \ref{qn} summarizes the explosion mechanism of the QN as well as the environment which leads to a SLSN. Section \ref{dsQN} examines the physics implemented in describing the interaction between ejecta of a SN and a QN.  Analysis of the effects of changing physical parameters on our model light curve is undertaken in section \ref{mod}.  In section \ref{results} we compare observations of eight SLSNe (SN 2005ap, SN 2006gy SN 2006tf, SN 2007bi, SN 2008es, SN 2008fz, PTF09cnd and PTF10cwr to QNe of different physical parameters.  A discussion of trends found fitting the SLSNe and spectral analysis for some targets is presented in section \ref{discuss}.  Finally our conclusions as well as predicted chemical signatures of our model are discussed in section \ref{conc}.

\begin{table*}

\caption{Comparison of proposed SLSNe models.}
 \label{compModelsTable}
 \begin{tabular}{@{}lccccc}
   & \vline & & Model & \\ 
  \hline              
 SLSNe Property & \vline  & PISN & CSM & Magnetar & dsQN \\    
\hline 
  Energy Mechanism & \vline  & $\gamma = 4/3$ instability$^{\rm a}$ & Binary merger$^{\rm b}$   &   Rotational energy$^{\rm c}$ & QN explosion $^{\rm d}$  \\                
   Radiation Mechanism & \vline & Collision of ejecta$^{\rm a}$ & CSM/SN interaction$^{\rm e}$ & Synchrotron$^{\rm c}$ & QN/SN interaction$^{\rm d}$ \\      
  Progenitor mass (M$_{\odot}$) & \vline & 313$^{\rm f}$ - 1250$^{\rm f}$ & 100+$^{\rm g}$  & 8$^{\rm h}$ - 25$^{\rm i}$ & 20$^{\rm j}$ - 40$^{\rm j}$ \\
      X-rays & \vline & suppressed$^{\rm k}$ & suppressed$^{\rm l}$  & not discussed & suppressed$^{\rm j}$ \\
    Hydrogen in Spectra & \vline  & unlikely$^{\rm m}$ & necessary$^{\rm n}$ & not discussed & likely$^{\rm j}$ \\
    Cause of long-lasting broad lines & \vline & not expected$^{\rm o}$ & velocity of SN$^{\rm g}$ & not expected$^{\rm c}$ & temperature of inner shell$^{\rm j}$ \\
    Late stage luminosity & \vline & radioactivity$^{\rm a}$ & opaque outer region of CSM$^{\rm p}$  & inner bubble$^{\rm c}$ & inner shell emission$^{\rm j}$  \\

\hline  
 \end{tabular}
 $^{\rm a}$\cite{woosley07}, 
  $^{\rm b}$\cite{chev12},
  $^{\rm c}$\cite{kasen10},
   $^{\rm d}$\cite{leahy08},
$^{\rm e}$\cite{chev82},  
$^{\rm f}$\cite{pan12}, 
 $^{\rm g}$\cite{smith10},
 $^{\rm h}$\cite{weidemann77}
 $^{\rm i}$\cite{fryer99},
 $^{\rm j}$\cite{ouyed12},
   $^{\rm k}$\cite{blinnikov08},
   $^{\rm l}$\cite{chev12b},
  $^{\rm m}$\cite{yungelson08},
 $^{\rm n}$\cite{quimby11},
 $^{\rm o}$\cite{kasen12},
 $^{\rm p}$\cite{chev11},

\end{table*}

\section{Observations}
\label{obs}

For this analysis we have chosen eight SLSNe (SN 2005ap, SN 2006gy, SN 2006tf, SN 2007bi, SN 2008es, SN 2008fz, PTF09cnd and PTF10cwr) to study in the context of the quark nova \citep{ouyed02}.  Each SLSN target along with the peak magnitude, class and proposed models are summarized in Table \ref{obsTable}.

The first SLSN ever observed was SN 2006gy \citep{quimby07} which peaked in absolute R band magnitude at approximately -22.  The spectrum of SN 2006gy is dominated by a broad H$\alpha$ emission line \citep{smith10} and while SN 2006gy was exceptionally bright in visible light the event was surprisingly quiet in X-rays \citep{smith07}. 

Spurred by the discovery of SN 2006gy, \cite{quimby07} found another extremely bright (-22.7 peak absolute R-band magnitude) SN 2005ap; which remains the brightest SLSN ever observed.  The spectrum of SN 2005ap shows broad spectral lines (H$\alpha$, C {\sevensize III}, N {\sevensize III}) and similar to SN 2006gy, SN 2005ap was quiet in X-rays \citep{quimby07}.  
 
SN 2006tf was discovered by \cite{quimby07b} who noted that the spectrum closely resembles that of SN 2006gy.  Observations of the light curve of SN 2006tf missed the leading edge of the SLSN and thus the actual peak magnitude is unclear, however the total radiated energy of the SLSN was at least $7 \times 10^{50}$ ergs \citep{smith08}.  \cite{smith07b} noted that the light curve of SN 2006tf is characterized by a very slow luminosity decay rate ($\sim0.01$ Mag. day$^{-1}$).  The spectra of SN 2006tf displays a strong H$\alpha$ emission line that remains broad over time \citep{smith08}.  H$\alpha$ and H$\beta$ show an interesting evolution in which the red-side of the profile remains constants while the blue-side emission wing becomes more prominent with time \citep{smith08}.  There exists a blue-side absorption feature seen in He {\sevensize I} $\lambda5876$ and O {\sevensize I} $\lambda7774$ that is of comparable width to the blue-side absorption feature of H$\alpha$ \citep{smith08}. 

\cite{galyam09} discovered the SLSN SN 2007bi and identified it as a type Ic SN, noting that there was no sign of a CSM interaction.  The light curve of SN 2007bi peaks at $\sim -21.3$ in the R band and displays a slow luminosity decay rate ($\sim 0.01$ Mag. day$^{-1}$) \citep{galyam09}.  As discussed by \cite{young10}, the relatively high host galaxy metalicity is inconsistent with a PISN explanation for the SLSN.  The slowly evolving spectra of SN 2007bi shows strong oxygen and iron lines \citep{young10}.

Discovered by \cite{gezari09}, the light curve of SN 2008es peaks at $\sim -22.2$ in the R band and shows a fast luminosity decay rate ($\sim 0.042$ Mag. day$^{-1}$).  The total radiated energy in UV and visible is in excess of $10^{51}$ ergs, however consistent with other SLSN, SN 2008es was quiet in X-rays \citep{gezari09}.  The spectrum of SN 2008es is dominated by broad features that lack the narrow and intermediate width line emission typically associated with a CSM interaction \citep{miller09}.  The spectral evolution of SN 2008es shows that the broad components of the spectral lines become more prominent over time \citep{miller09}.  

SN 2008fz was discovered by \cite{drake10} who found the light curve to peak at $\sim -22.3$ in the V band and shows a similar slow evolution to that of SN 2006gy.  The spectrum of SN 2008fz displays strong Balmer lines that are initially narrow but become broad over time, \cite{drake10} noted that the H$\alpha$ emission line of SN 2008fz is similar to that of SN 2006gy for the same epoch.

We chose to study PTF09cnd and PTF10cwr as representative members of the new hydrogen-poor SLSN class \citep{quimby11}.  These SLSNe were selected because they display the two extremes of light curve morphology for this class, PTF09cnd has the brightest and broadest u-band light curve in the class and PTF10cwr the dimmest and narrowest.  
 The luminosity decay rate of both PTF09cnd and PTF10cwr are inconsistent with radioactive decay and the spectrum shows no signs of a CSM interaction \citep{quimby11}.  PTF10cwr was also observed in the B band by \cite{pastorello10}, in which it peaks at $\sim -21.2$, and alternatively is referred to as SN 2010gx.  As mentioned in the supplemental material associated with \cite{quimby11}, the SLSNe PTF09cnd and PTF10cwr are quiet in X-rays relative to the amount of energy radiated in the visual bands.

\section{Quark Nova}
\label{qn}

\subsection{Explosive Mechanism}
The existence of up-down-strange (UDS) quark matter as a state of matter more stable than hadronic has been theorized for several decades \citep{itoh70,bodmer71,witten84}.  However, \citet{ouyed02} was first to propose that if a phase transition to UDS matter were to occur at the core of a neutron star the explosive result would be capable of ejecting the outer layer of the neutron star leaving behind a quark star.  This violent new pathway for massive star evolution is called a QN.  It has been shown for massive neutron stars that either the accretion of SN fall-back material or the spin-down evolution can cause the core density to exceed that of quark deconfinement, triggering a QN \citep{staff06}.  Detailed study of hadronic to UDS burning suggested that turbulence in the conversion front enhances the burning speed to sonic or supersonic velocities which in turn would make the transition from neutron star to quark star detonative \citep{nieb10,ouyed11b}.  Further investigation of neutrino and photon emission from a QN found that 10$^{-5}$ - 10$^{-3}$ M$_{\sun}$ could be ejected with up to 10$^{53}$ erg of kinetic energy, which means ejecta with a Lorentz factor of 10-100 \citep{vogt04,ouyed09}.

\subsection{Timing}
The length of time between SN and QN explosion ($t_{\rm delay}$) plays a crucial role in determining the subsequent evolution of the star.  When the delay is short ($\lesssim 8$ days) the SN envelope is still dense and the energy of the impacting QN ejecta is used up in spallation of the inner region of the SN envelope \citep{ouyed11}.  This leads to the destruction of $^{56}$Ni and the formation of $^{44}$Ti and results in a subluminous SN. The reduced luminosity is due to the lack of radioactive decay of $^{56}$Ni \citep{ouyed11}.  When the time delay is long (on the order of months or longer) the SN envelope will have become too diffuse to significantly interact with the QN ejecta.  For massive QN ejecta with long time delays fall-back of the ejected material can occur which has implications for $\gamma$-ray bursts, soft $\gamma$-ray repeaters and anomolous X-ray pulsars \citep{ouyed07a,ouyed07b,koning12}.

This work will focus on the scenario in which a SN is followed on the order of weeks by a QN.  In this case, referred to as a dual-shock QN (dsQN), the expanded SN envelope is bombarded by the QN ejecta reheating the SN envelope.  The extended size and high temperature of the re-shocked SN envelope yields a brilliant radiance capable of reproducing the observed luminous peaks of SLSNe.  As the radiation from the extended envelope fades, an inner region of mixed QN and SN material is revealed which can explain both the luminosity decay rate of SLSNe as well as curious spectral features. 

\section{Dual-Shock Quark Nova}
\label{dsQN}

The evolution of a dsQN can be considered as three distinct phases; \textit{delay, shock} and \textit{cooling}.  During the \textit{delay} phase the SN envelope expands homologously ($v \propto r$) while the neutron star evolves towards a QN explosion.  For fiducial values the radius of the SN envelope will reach $\sim 10^{15}$ cm during this phase.  The end of the \textit{delay phase} is marked by the detonation of the QN. 

 At the beginning of the \textit{shock} phase the ultra-relativistic QN ejecta \citep{keranen05,ouyed09a} quickly catches and slams into the inner edge of the SN envelope.  This collision creates a shock front that propels through the SN envelope reheating it to a temperature $\sim 10^9$ K \citep{leahy08}.  As the shock progresses the inner region of the re-shocked SN envelope mixes with the impacting QN ejecta to create a thin shell interior to the envelope (referred to as the hot plate in paper I).  

The end of the \textit{shock} phase and beginning of the \textit{cooling} phase is defined to be the moment when the shock breaks out of the SN envelope.  By this time the inner shell will be fully formed and slowly coasting inside of the shocked envelope.  During the \textit{cooling} phase both the envelope and the inner shell will cool via adiabatic expansion and radiative emission.  Following the same methodology used in \cite{ouyed12} (furthermore to be referred to as paper I), for the work presented here we model the light curves of dsQNe during their \textit{cooling} phase.

%

%

\subsection{The Inner Shell}
\label{shell}

As the QN ejecta ploughs through the SN envelope material builds up to form the inner shell.  The mass ($M_{\rm sh}$) and velocity ($v_{\rm sh}$) of the inner shell are found to be strongly correlated through conservation of energy and momentum.  Processes such as turbulent mixing and reverse shocks cause the thickness ($\Delta R_{\rm sh}$) of the inner shell to remain constant once it has begun to coast (defined to begin at radius, $R_{\rm sh, 0}$).  A detailed description of the formation of the inner shell as well as emission methodology used in this work can be found in paper I.  The velocity of the inner shell that we found as a fit parameter in paper I was approximately 10\% of our theoretically predicted value.  Upon revisiting our derivation of R band luminosity used in paper I, we found that there was a missing factor of $\pi$ and with this small correction the inner shell velocity used in this work is now in full agreement with the theoretical value (see eqn. 7 of paper I).

The inner shell is fully parameterized by: temperature, mass, velocity, thickness and coasting radius. The mass and velocity are determined through conservation of energy and momentum, the derivation of the formulae used to describe these parameters can be found in paper I (equations 6 and 7 respectively).   For our analysis of each SLSN target considered in this work we use the same coasting radius for the inner shell, namely the best fit value of $R_{\rm sh, 0} = 4 \times 10^{14}$ cm that was found in paper I.  The only parameter pertaining solely to the inner shell that we allow to be adjusted is shell thickness.  

A detailed look at the formation of the inner shell using a full hydrodynamic treatment would help to understand how physical parameters of the inner shell are affected by changing initial conditions such as the time delay between SN and QN.

\subsection{The Envelope}
\label{env}

Fiducial values, found in paper I, for the shock speed ($v_{\rm shock} = 6000$ km s$^{-1}$) and the outer edge velocity of the homologously expanding ($v\propto r$) envelope ($v_{\rm SN, max} = 4100$ km s$^{-1}$) are used for the work presented here.  By fixing $v_{\rm shock}$ and $v_{\rm SN, max}$ the variation in the time at which the \textit{cooling} phase begins is uniquely determined by $t_{\rm delay}$.  Following the same methodology as paper I, a simple temperature profile is considered for the envelope where $T(r) \propto r ^{-\beta}$ with $\beta = 0.2$.  In this work the progenitor star mass ($M_{\star}$) is a free parameter and the mass of the inner shell ($M_{\rm sh}$) is prescribed by conservation laws, thus the mass of the envelope is found simply as $M_{\rm env} = M_{\star}-M_{\rm sh}$.

\subsection{Cooling}
\label{cool}
Since the inner shell and the envelope are initially both parts of the same physical structure, namely the re-shocked SN envelope, they both start with same initial temperature ($T_0$).  As the envelope and the inner shell expand they cool adiabatically, due to the difference in geometries (the envelope expanding spherically and the inner shell with constant thickness) they follow different cooling profiles. The temperature of the envelope evolves as $ T_{\rm env}(t) \propto T_0\,t^{-2}$ and for the inner shell $ T_{\rm sh}(t) \propto T_0\,t^{-4/3}$ (see paper I for derivation). 

As continuum radiation is emitted by the envelope an equivalent amount of thermal energy is removed, thereby conserving energy.  Since the \textit{cooling} phase begins when the envelope is still relatively optically thick, radiative cooling starts at the outer edge of the envelope.  Due to radiative cooling, over time a hot-cold interface progresses inward through the envelope.  In our model once an outer layer of the envelope is radiatively cooled it no longer emits radiation.  At the late stages of the dsQN light curve, radiative cooling will have caused the hot-cold interface to progress such that most of the envelope has been cooled.  By this time the envelope has also expanded causing it to become more optically thin and thus making our optically thick radiative cooling approximation less valid.  In reality as the envelope becomes more diffuse radiative cooling will cause the whole envelope to cool rather than just the outer layers and lead to an increase in the cooling rate of the envelope with time.  The effect of this can be seen in the envelope contribution to the dsQN light curve (plotted as a blue dotted line in Fig. \ref{genericLC}): as the change in slope around day 170 is due to the envelope remaining hotter than physically likely.  Since the envelope is optically thin when this cooling artefact becomes significant, in most cases luminosity from the inner shell will already dominate the overall light curve rendering the problem moot.  Rather than adding a more complicated radiative cooling law to our model we will simply note when the effect is seen during our fits to observations.

\begin{table}

\caption{List of SLSNe targets.}
\label{obsTable}
\begin{tabular*}{\columnwidth}{@{}llll}
\hline
 Name & Type & Peak Magnitude & Proposed Models  \\
SN 2005ap & H-poor SLSN$^{\rm a}$ & -22.7$^{\rm b}$ & CSM$^{\rm a}$, PISN$^{\rm a}$,dsQN$^{\rm c}$ \\
SN 2006gy & IIn$^{\rm d}$  & -22 $^{\rm e}$ & PISN$^{\rm f}$, CSM$^{\rm e}$, dsQN$^{\rm c,g}$ \\
SN 2006tf & IIn$^{\rm h}$  & $\sim$-20.8$^{\rm h}$ & CSM$^{\rm h}$ \\
SN 2007bi& Ic$^{\rm i}$  & -21.3$^{\rm i}$ & CC$^{\rm j}$, PISN$^{\rm i}$\\
SN 2008es & II-L$^{\rm k}$  & -22.2$^{\rm k}$ & CSM$^{\rm k}$\\
SN 2008fz & IIn$^{\rm l}$  & -22.3$^{\rm l}$ & none$^{\rm l}$ \\
PTF09cnd & H-poor SLSN$^{\rm a}$  & -22$^{\rm a}$ & CSM$^{\rm a}$, PISN$^{\rm a}$ \\
PTF10cwr & H-poor SLSN$^{\rm a}$  & -21.2$^{\rm m}$ & CSM$^{\rm a}$, PISN$^{\rm a}$ \\
\hline
\end{tabular*}
\footnotesize{
$^{\rm a}$ \cite{quimby11},
$^{\rm b}$ \cite{quimby07},
$^{\rm c}$ \cite{leahy08},
$^{\rm d}$ \cite{quimby05},
$^{\rm e}$ \cite{smith07},
$^{\rm f}$ \cite{woosley07},
$^{\rm g}$ \cite{ouyed12},
$^{\rm h}$ \cite{smith08},
$^{\rm i}$ \cite{galyam09},
$^{\rm j}$ \cite{young10},
$^{\rm k}$ \cite{gezari09},
$^{\rm l}$ \cite{drake10},
$^{\rm m}$ \cite{pastorello10}

}
\end{table}

\section{Model}
\label{mod}
For this work we use the astrophysical modelling software {\sevensize SHAPE}, which allows us to construct the 3-D geometry of the dsQN scenario and perform radiative transfer \citep{steffen10}.  The temperature as well as the dimensions of the geometry are governed by the physics described in section \ref{dsQN}.

\subsection{Radiative Transfer Parameters}
\label{params}


For this analysis the radiative transfer calculation follows the same methodology used in paper I in which for each frequency ($\nu$) an emission and absorption coefficient are specified.  The emission coefficient used for the envelope has the form

\begin{equation}
\label{emi}
j_{\nu} = \frac{A\,n_{\rm e}^2}{T^{3/2}}e^{h\nu / kT}
\end{equation}

\noindent
where $A$ is a multiplicative factor which is related to the underlying radiative process.  For this analysis we chose to fix $A$ over the filter passband of each studied SLSN.  For the R band $A =  5 \times 10^5$, V band $A =  1 \times 10^6$,  B band $A =  7 \times 10^4$,  u band $A =  5 \times 10^5$. The variations required in $A$ allude to an emission mechanism more complex than our approximation described by eqn. \ref{emi}.  Assuming local thermodynamic equilibrium the absorption coefficient corresponding to our $j_{\nu}$ is found by means of the Planck function ($B_{\nu}$) to be

\begin{equation}
\label{abs}
k_{\nu} = \frac{j_{\nu}}{B_{\nu}}
\end{equation}
\noindent
We as well include a Thompson scattering term to equation \ref{abs} of the form

\begin{equation}
\label{scat}
k_{\nu , {\rm TH}} = B\, n_{\rm e} \, \sigma_{\rm TH}
\end{equation}
\noindent
where $\sigma_{\rm TH}$ is the Thompson scattering cross-section and $B$ is a multiplicative factor which represents the fraction of scattered light that is not scattered back into the beam.  For this work we set $B = 5 \times 10 ^{-4}$, which is the same as that used in paper I.  Further details on {\sevensize SHAPE} and the radiative transfer calculation can be found in paper I. 

\subsection{Fiducial Model Characteristics}
\label{generic}

\begin{figure*}

\includegraphics[scale=1]{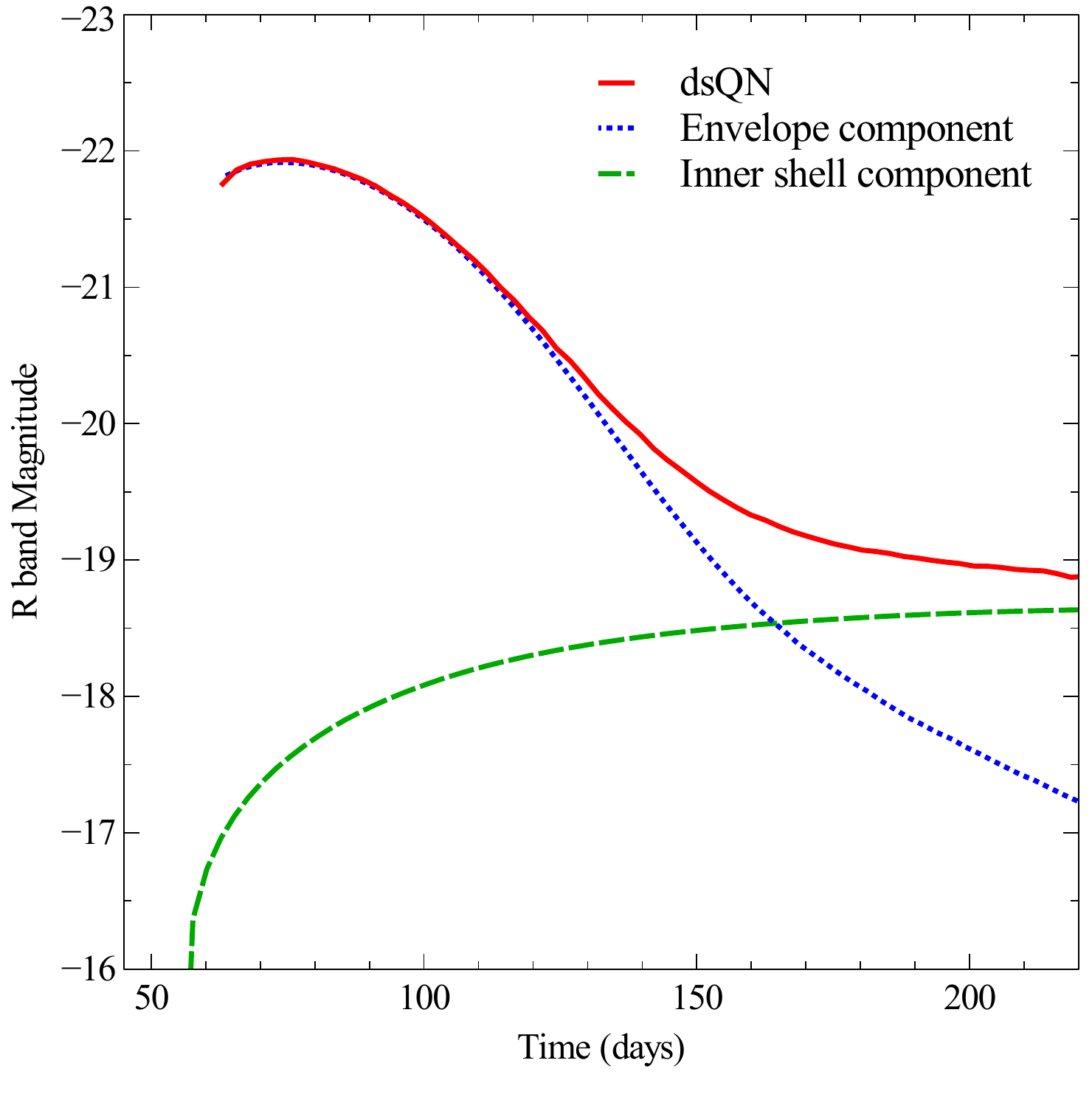}
\caption{Plotted as a red solid line is the R-band light curve of the dsQN.  Radiation from the two components of the dsQN are also plotted.  Radiation from the envelope is represented by the blue dashed line and that from the inner shell is denoted by the green dotted line.}
\label{genericLC}
\end{figure*}

Before fitting the observed SLSNe light curves with the dsQN model we felt it would be informative to first explore the light curves of dsQNe in a more general sense.  To this end we built a generic dsQN model with the following physical parameters: $M_{\star} = 25 M_{\sun}$, $T_0 = 2.5 \times 10^9$ K, $\Delta R_{\rm sh} = 3 \times 10^{13}$cm and $t_{\rm delay} = 15$ days.  The light curve associated with this model is plotted as a red solid line in Fig. \ref{genericLC}.  The component of the light curve caused by emission from the envelope is plotted as a blue dotted line in Fig. \ref{genericLC} and the contribution from the inner shell is represented by the green dashed line.  As can be seen in Fig. \ref{genericLC} the broad peak in the dsQN light curve is due to radiation emitted from the envelope.  As time progresses the envelope cools and becomes less dense causing a rapid drop in emission, this allows for radiation from the slowly coasting and thus slowly cooling inner shell to begin to shine through.  The post-peak luminosity decay rate in the overall light curve of the dsQN is slowed due to emission from the inner shell.  In the late stages of the dsQN light curve there can exist a plateau due to emission from the inner shell.

\begin{figure*}

\includegraphics[scale=1.2]{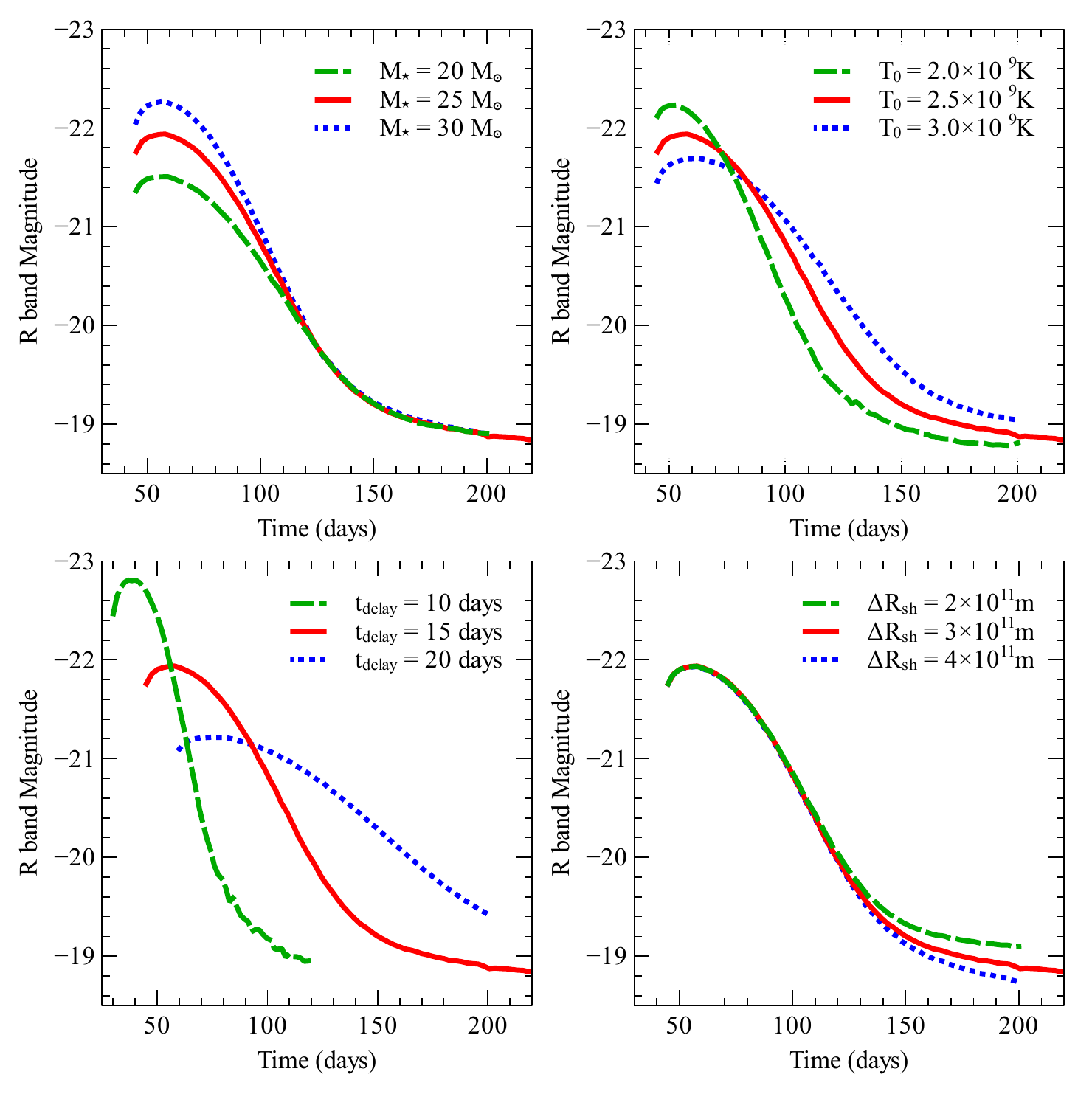}

\caption{In each panel the overlay of the R-band light curve of dsQN models with three different sets of physical parameters is plotted.  For each panel one physical parameter is varied while the remaining parameters are held constant (values given in text). \textit{Top-left: Mass of envelope ($M_{\star}$).} $M_{\star} = 20 M_{\sun}$  (green dashed line), $M_{\star} = 25 M_{\sun}$ (red solid line) and $M_{\star} = 30 M_{\sun}$ (blue dotted line). \textit{Top-right: Initial shock temperature ($T_0$).} $T_0 = 2 \times 10^9$ K (green dashed line), $T_0 = 2.5 \times 10^9$ K (red solid line) and $T_0 = 3 \times 10^9$ K (blue dotted line).  \textit{Bottom-left: Time delay between SN and QN ($t_{\rm delay}$).}  $t_{\rm delay} = 10$ days (green dashed line), $t_{\rm delay} = 15$ days (red solid line) and $t_{\rm delay} = 20$ days (blue dotted line).  \textit{Bottom-right: Inner shell thickness (${\Delta}R_{\rm sh}$).}  ${\Delta}R_{\rm sh} = 2 \times 10^{11}$m (green dashed line), ${\Delta}R_{\rm sh} = 3 \times 10^{11}$m (red solid line) and ${\Delta}R_{\rm sh} = 4 \times 10^{11}$m (blue dotted line).   } 
\label{varyConstA}
\end{figure*}

In this section we will investigate the effect on light curve morphology of varying the physical parameters of a dsQN.  For comparative purposes in each of the panels in Fig. \ref{varyConstA} the red solid line denotes the same light curve as the overall light curve seen in Fig. \ref{genericLC} (red solid line).  Then in each panel of Fig. \ref{varyConstA} one physical parameter is adjusted to show its effect on the light curve, the higher value is denoted by the blue dotted line and lower value by the green dashed line.  Clock-wise starting from the upper-left panel of Fig. \ref{varyConstA}; varying $M_{\star}$ ($20 M_{\sun}$, $25 M_{\sun}$ and $30 M_{\sun}$), varying $T_{\rm 0}$ ($2\times10^9$K, $2.5\times10^9$K and $3\times10^9$K), varying ${\Delta}R_{\rm sh}$ ($2\times10^{13}$cm, $3\times10^{13}$cm and $4\times10^{13}$cm) and varying $t_{\rm delay}$ (10 days, 15 days and 20 days).  

As seen in the upper-left panel of Fig. \ref{varyConstA} the effect of varying $M_{\star}$ is simply scaling the height of the peak, with higher mass yielding a higher peak.  As the post-peak luminosity drops support by radiation from the underlying inner shell kicks in at the same time for each of the light curves.  This convergence is caused by the fact that changing the mass of the envelope has no effect on the emission from the inner shell. 

 Changing $T_{\rm 0}$ of the dsQN (as seen in the upper-right panel of Fig. \ref{varyConstA}) as well acts to scale the height of the peak, however the height of the peak is inversely proportional to $T_{\rm 0}$ (due to the form of the emission coefficient, see eqn. \ref{emi}).  In the case of varying $T_{\rm 0}$ the support from the inner shell on the overall light curve begins at different times because temperature directly effects the amount of radiation that can be emitted by the inner shell.  For higher $T_0$ the luminosity of the inner shell is greater and its effect on the overall light curve can be seen sooner. 
 
  As shown in Fig. \ref{genericLC} the effect on the dsQN light curve of radiation from the inner shell can only be seen in the late stages once emission from the envelope has faded sufficiently.  Thus as expected the lower-right panel of Fig. \ref{varyConstA} shows that changing ${\Delta}R_{\rm sh}$ only impacts the tail of the dsQN light curve.  All things being equal, a thinner shell implies higher density and thus more intense emission from the inner shell.   The effect of increased density causing increased radiation can be seen in the lower-right panel on Fig. \ref{varyConstA} where the thinner ${\Delta}R_{\rm sh}$ light curve has a more prominent plateau.  
  
  The physical parameter that has the most dramatic effect on light curve morphology is $t_{\rm delay}$, which can be seen in the lower-left panel of Fig. \ref{varyConstA}.  A shorter $t_{\rm delay}$ implies that the SN envelope is more dense when it is re-shocked which translates to higher peak in the light curve.  The shorter $t_{\rm delay}$ also demands that the SN envelope is smaller when it is re-shocked, leading to significant adiabatic losses and a steep luminosity decay rate.  The dsQN with a longer $t_{\rm delay}$ can not achieve as high of a luminosity peak, however reduced adiabatic losses allow the light curve to remain at a higher relative brightness for a longer time when compared to a dsQN with a shorter $t_{\rm delay}$.  Another effect of adjusting $t_{\rm delay}$ is to change the time of peak luminosity relative to the time of SN explosion.  A longer $t_{\rm delay}$ dsQN will peak in brightness later in time due to the simple fact that the shock must traverse a larger envelope before it can break out.
  
\begin{figure*}

\includegraphics[scale=1.2]{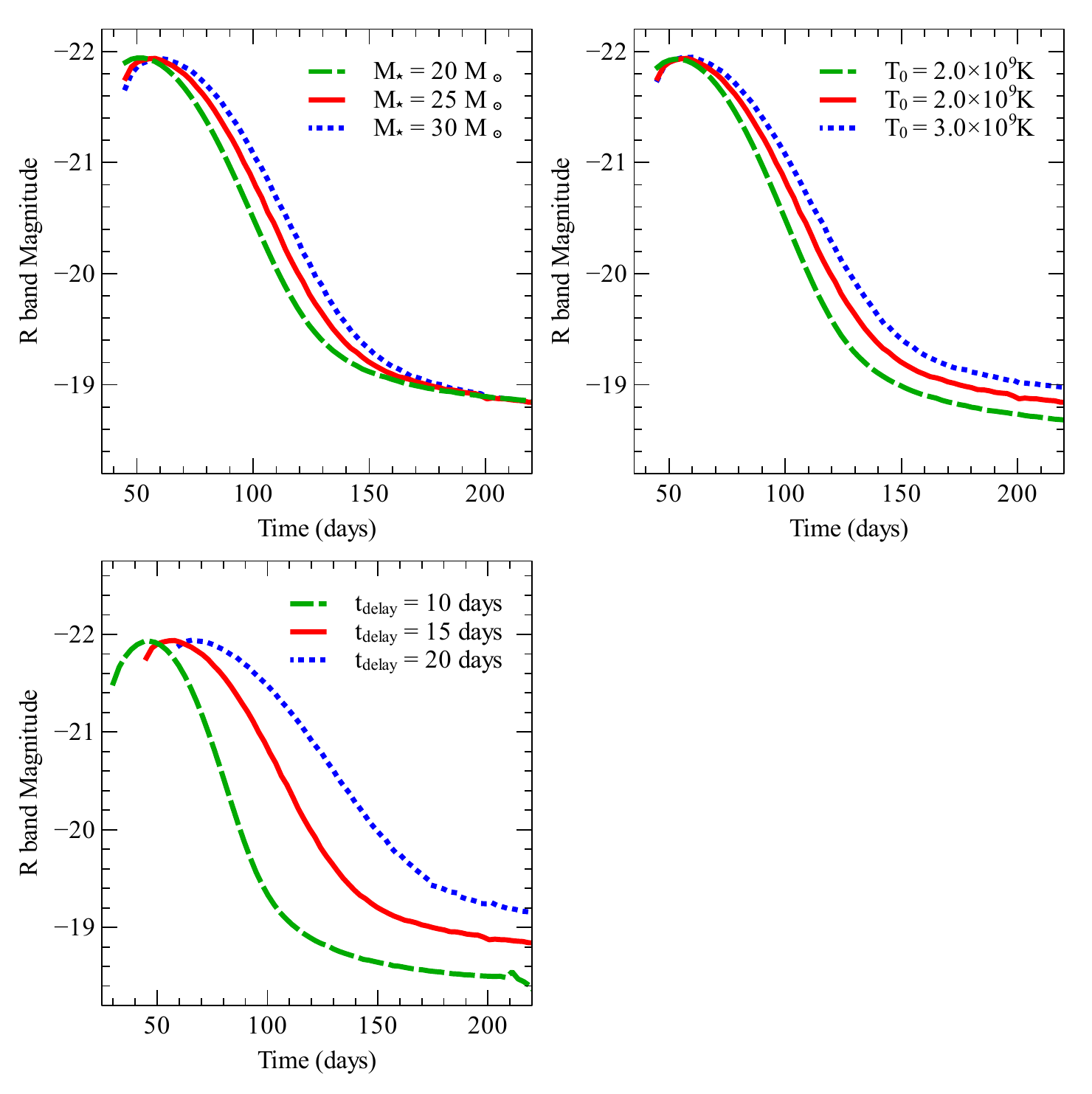}
\caption{In each panel the overlay of the R-band light curve of dsQN models with three different sets of physical parameters is plotted.   Each light curve is scaled using the radiative transfer parameter $A$ such that the peaks reach the same absolute magnitude.  For each panel one physical parameter is varied while the remaining parameters are held constant (values given in text).  \textit{Top-left: Mass of envelope ($M_{\star}$).} $M_{\star} = 20 M_{\sun}$  (green dashed line), $M_{\star} = 25 M_{\sun}$ (red solid line) and $M_{\star} = 30 M_{\sun}$ (blue dotted line). \textit{Top-right: Initial shock temperature ($T_0$).} $T_0 = 2 \times 10^9$ K (green dashed line), $T_0 = 2.5 \times 10^9$ K (red solid line) and $T_0 = 3 \times 10^9$ K (blue dotted line).  \textit{Bottom-left: Time delay between SN and QN ($t_{\rm delay}$).}  $t_{\rm delay} = 10$ days (green dashed line), $t_{\rm delay} = 15$ days (red solid line) and $t_{\rm delay} = 20$ days (blue dotted line). } 
\label{varyA}
\end{figure*}

From Fig. \ref{varyConstA} it is clear that changing the value of $M_{\star}$, $T_0$ or $t_{\rm delay}$ all effect the height of the luminosity peak of the light curve of a dsQN.  Our approximated emission coefficient (see eqn. \ref{emi}) contains the free parameter $A$ which as well directly effects the amount of radiation emitted in our model at any given time.  Thus we can use $A$ to gain a better visualization of the effect of changing each physical parameter ($M_{\star}$, $T_0$ and $t_{\rm delay}$) on the shape of the light curve.  In Fig. \ref{varyA} we perform the same comparison as shown in Fig. \ref{varyConstA} with the addition that each light curve is rescaled using $A$ such that the absolute magnitude of the light curve in the R band peaks at -22.  By comparing the light curves in the upper two panels of Fig. \ref{varyA} it is clear that neither changing $M_{\star}$ nor $T_0$ significantly affect the post-peak luminosity decay rate of the dsQN.  As seen in the upper-left panel of Fig. \ref{varyA} increasing the mass of the envelope marginally increases the breadth of the light curve peak but does not affect the slope.  Turning to the upper-right panel of Fig. \ref{varyA} one can see that increasing $T_0$ has the same effect on the morphology of the light curve peak as increasing $M_{\star}$.  The only difference between adjusting $M_{\star}$ and $T_0$ is that the intensity of the radiation from the inner shell increases with $T_0$ causing the plateau to vary directly with $T_0$. 

From Figures \ref{varyConstA} and \ref{varyA} we can see that varying $T_0$ and $M_{\star}$ effectively only act as scaling factors, with minimal effect on the luminosity decay rate of the light curve.  Since the effect is virtually the same, for all our fits to observed SLSN light curves we choose to fix $T_0$ at $2.5 \times 10^9$K and allow $M_{\star}$ to vary.  The lower-left panel of Fig. \ref{varyA} displays the effect of varying the delay between SN and QN.  As was shown in Fig. \ref{varyConstA} increasing $t_{\rm delay}$ causes the peak luminosity to occur later with respect to the time of SN explosion.  Another consequence of increasing $t_{\rm delay}$ is to slow down the luminosity decay rate of the post-peak light curve of a dsQN.  As we have found in this analysis the only parameter that we can adjust in order to change the luminosity decay rate in the dsQN model is $t_{\rm delay}$.

\section{Fits to Observations}
\label{results}

\begin{center}
\begin{figure*}
\includegraphics[scale=0.6]{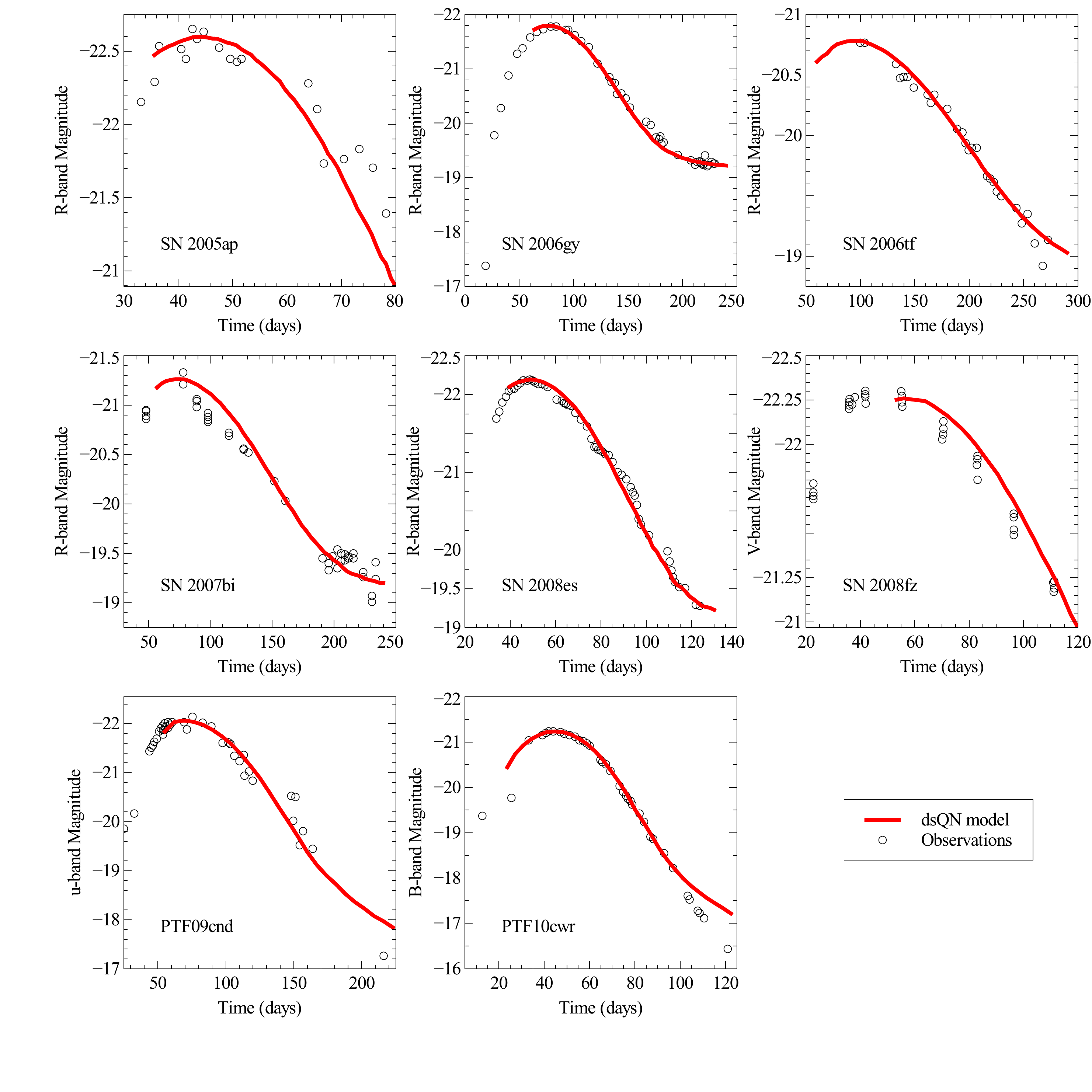}
\caption{For each panel the observed light curve is plotted with open black circles and the dsQN model is plotted as a solid red line.  Time since the inferred SN explosion is plotted along the horizontal axis and the absolute magnitude in the observed band is plotted on the vertical axis.  The best fit parameters used to generate these fits can be found in table \ref{paramTable}.  From left-to-right and top-to-bottom the panels represent: SN 2005ap, SN 2006gy, SN 2006tf, SN 2007bi, SN 2008es, SN 2008fz, PTF09cnd and PTF10cwr.} 
\label{fits}
\end{figure*}
\end{center}

\begin{figure*}
\includegraphics[scale=1]{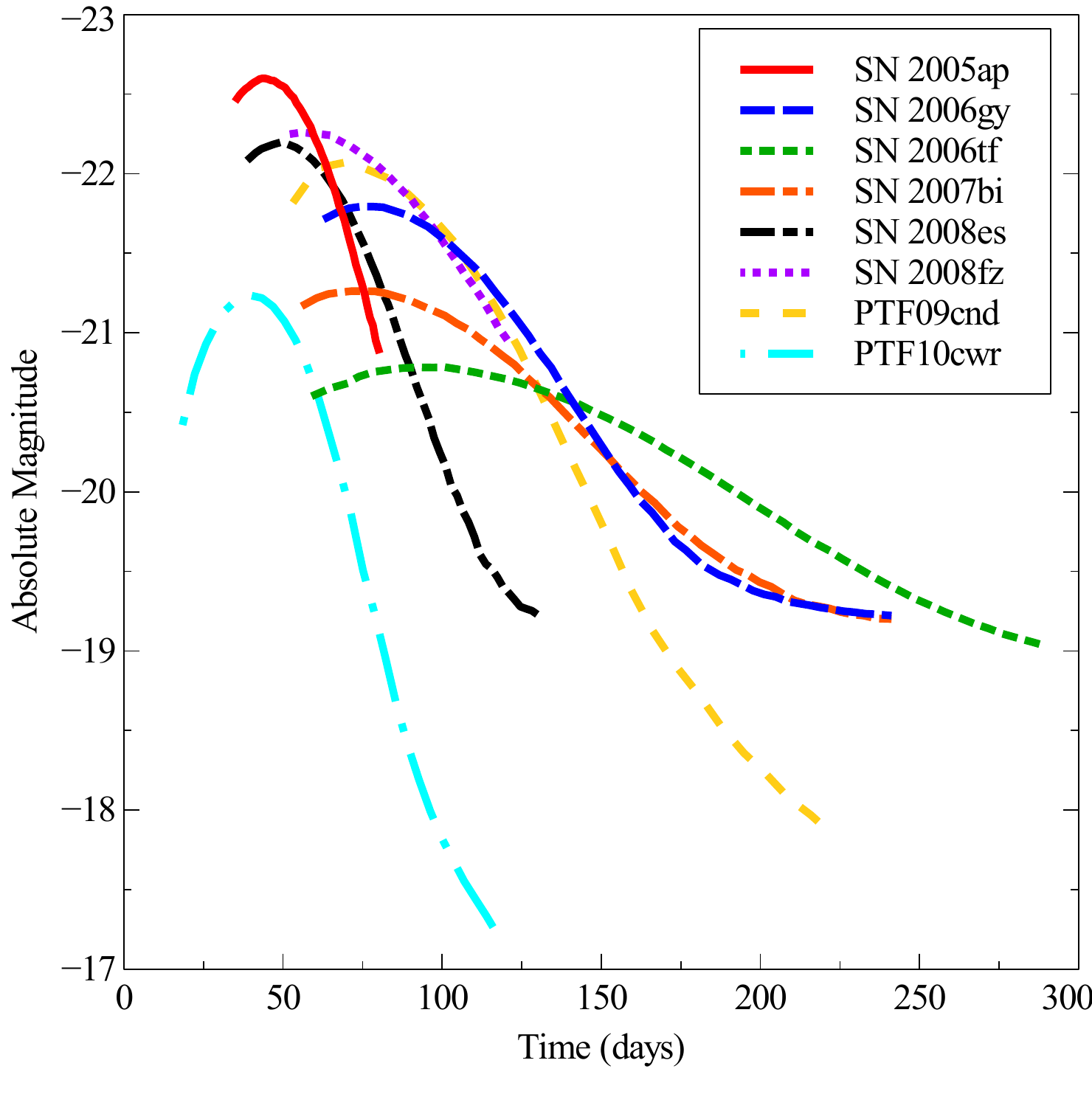}
\caption{The best fit dsQN model light curve for each of the SLSNe studied in this work are plotted on the same axis.} 
\label{fits_over}
\end{figure*}

For this work we have fit the light curve of eight SLSNe using the dsQN model.  The parameters that we adjusted to fit each set of observations were; $t_{\rm delay}$, $\Delta R_{\rm sh}$ and $M_{\star}$.  Of these three free parameters $t_{\rm delay}$ and $M_{\star}$ only significantly affect radiation from the envelope while $\Delta R_{\rm sh}$ only affects emission from the inner shell.  As was shown in section \ref{generic}, varying $M_{\star}$ or $T_0$ have the same effect on the dsQN light curve.  Thus for simplicity  we fixed $T_0 = 2.5 \times 10^9$ K and allowed $M_{\star}$ to vary between model fits.   The best fit values of these parameters for each SLSNe surveyed can be found in Table \ref{paramTable}.  For each panel of Fig. \ref{fits} the observations are plotted with open circles and the best fit dsQN light curve is plotted as a red solid line.  Fig. \ref{fits_over} displays each of the best fit dsQN light curves plotted on the same axes.  From Fig. \ref{fits_over} it can be seen that for dsQNe in which an inner shell is formed (all but PTF09cnd and PTF10cwr) the magnitude of the radiation from the inner shell is remarkably consistent.

\subsection{SN 2005ap}
 \label{05apFit}
The best fit progenitor mass for SN 2005ap was $28 M_{\sun}$.  The high peak absolute magnitude and fast luminosity decay rate of SN 2005ap required a short time delay ($t_{\rm delay} = 11.25$ days) dsQN to fit the observations (see top row left panel of Fig. \ref{fits}).  Since the observations of SN 2005ap were only carried out for approximately 35 days post-peak, the luminosity had not dropped enough to potentially see contribution from the inner shell.  Thus we were unable to determine if an inner shell was present in the dsQN model of this SLSN.

\subsection{SN 2006gy}
 \label{06gyFit} 

In the proof-of-principle paper \citep{ouyed12} the fits to SN 2006gy ignored a $\pi$ factor in the R band magnitude calculation, the fits presented here take this into account and thus are more accurate.  Our fit to the observed light curve of SN 2006gy is displayed in the top row middle panel of Fig. \ref{fits}.  A $30 M_{\sun}$ progenitor star was used in our model and the broad peak of SN 200gy required a time delay of 17 days.  The distinct plateau in the late stages of the light curve was fit using a inner shell $2 \times 10^{13}$cm thick.

\subsection{SN 2006tf}
 \label{06tfFit} 
Although the observations of SN 2006tf do not include the leading edge of the light curve, the slow luminosity decay rate demanded that the dsQN used in this fit have the longest time delay of all SLSNe modelled in this work ($t_{\rm delay} = 22.5$ days).  A progenitor star of $26 M_{\sun}$ was used in our fit and the late stages of the light curve is affected by contribution from an $4\times 10^{13}$cm thick inner shell.  Our fit to the observed light curve of SN 2006tf is found in the top row right panel of Fig. \ref{fits}.

\subsection{SN 2007bi}
 \label{07biFit} 
A plot of our fit to the light curve of SN 2007bi is displayed in the middle row left panel of Fig. \ref{fits}.  A progenitor star of $25.5 M _{\sun}$ was used in our model and the delay between SN and QN that best fit the shape of the SN 2007bi light curve was $t_{\rm delay} = 18.5$ days.  The late stage plateau of the light curve was fit in our dsQN model by radiation predominantly from a $2.5\times 10^{13}$cm thick inner shell.

\subsection{SN 2008es}
 \label{08esFit}
The narrow peak of the light curve of SN 2008es was fit by dsQN of a $28 M _{\sun}$ progenitor star with a time delay between SN and QN of 12.5 days.  At approximately 100 days after SN explosion the light curve of SN 2008es begins to be supported by luminosity originating from a $3\times 10^{13}$cm thick inner shell.  A plot of our fit to the light curve of SN 2008es is displayed in the middle row middle panel of Fig. \ref{fits}.

\subsection{SN 2008fz}
 \label{08fzFit}
The combination of high peak luminosity and broad light curve made SN 2008fz require a $35 M_{\sun}$ progenitor star and time delay of 20 days between SN and QN.  Our dsQN fit to the light curve is found in the middle row middle panel of Fig. \ref{fits}.  The late stages of the SN 2008fz light curve is supported by luminosity from a $5\times 10^{12}$cm inner shell.

\subsection{PTF09cnd}
 \label{09cndFit}
A plot of our fit to the light curve of PTF09cnd is displayed in the bottom row left panel of Fig. \ref{fits}.  The best fit parameters for PTF09cnd ($t_{\rm delay} = 17$ days and $M_{\star} = 31 M_{\sun}$) are similar to that of SN 2006gy with one exception.  For PTF09cnd we found that the inner shell had to be nearly 10 times thicker than that of SN 2006gy ($2\times 10^{14}$ cm, or approximately two fifths of the initial radius of the inner shell) in order to fit the late stages of the light curve.  This difference in inner shell geometry may be attributed to difference in the abundance of the inner SN envelope and higher order mixing effects related to specific dynamics of the QN/SN interaction.  Another possible explanation for the reduced radiation from the inner shell at the late stages of the light curve of PTF09cnd is that the inner shell is cooling faster than the imposed $T\propto t^{-4/3}$ law.  A larger cooling index would simply imply that the shell is expanding in thickness with time.  A more detailed study of the hydrodynamics of the QN/SN interaction would help to further understand the formation process of the inner shell.  The late stages of our dsQN model fit to PTF09cnd displays a change in slope which is an artefact of the radiative cooling approximation discussed in section \ref{cool}.

\subsection{PTF10cwr / SN 2010gx}
 \label{10gxFit} 
Displayed in the bottom row middle panel of Fig. \ref{fits} is the dsQN model fit to the observed light curve of PTF10cwr.  This SLSN displayed the narrowest light curve of all those studied in this work and thus was fit with the shortest time delay (7.5 days) dsQN.  A $25 M_{\sun}$ progenitor star was used in our best fit model and the light curve of PTF10cwr showed no sign of the presence of an inner shell.  The late stages of our dsQN model fit to PTF10cwr displays a change in slope which is an artefact of the radiative cooling approximation discussed in section \ref{cool}.

\begin{table*}

\caption{Comparison of dsQN model parameters used to fit each SLSNe.}
\label{paramTable}
\begin{tabular}{@{}lccc}
SLSNe &  Time Delay ($t_{\rm delay}$)  & Mass ( $M_{\star}$) & Shell Thickness ($\Delta R_{\rm sh}$)  \\
 & [days] & [$M_{\sun}$]  & [$10^{13}$ cm] \\
\hline
SN 2005ap & 11.25 & 28 & N/A \\
SN 2006gy & 17.0 & 30 & 2.0 \\
SN 2006tf & 22.5 & 26 & 4.0 \\
SN 2007bi & 18.5 & 25.5 & 2.5 \\
SN 2008es & 12.5 & 28 & 3.0 \\
SN 2008fz & 20.0 & 35 & 0.5 \\
PTF09cnd & 17.0 & 31 & $\gtrsim$ 20  \\
PTF10cwr & 7.5 & 25 & N/A \\
\hline
\end{tabular}
\end{table*}

\section{Discussion}
\label{discuss}

\subsection{Spectra}
 \label{spectraDisc}

\begin{figure*}

\includegraphics[scale=0.66]{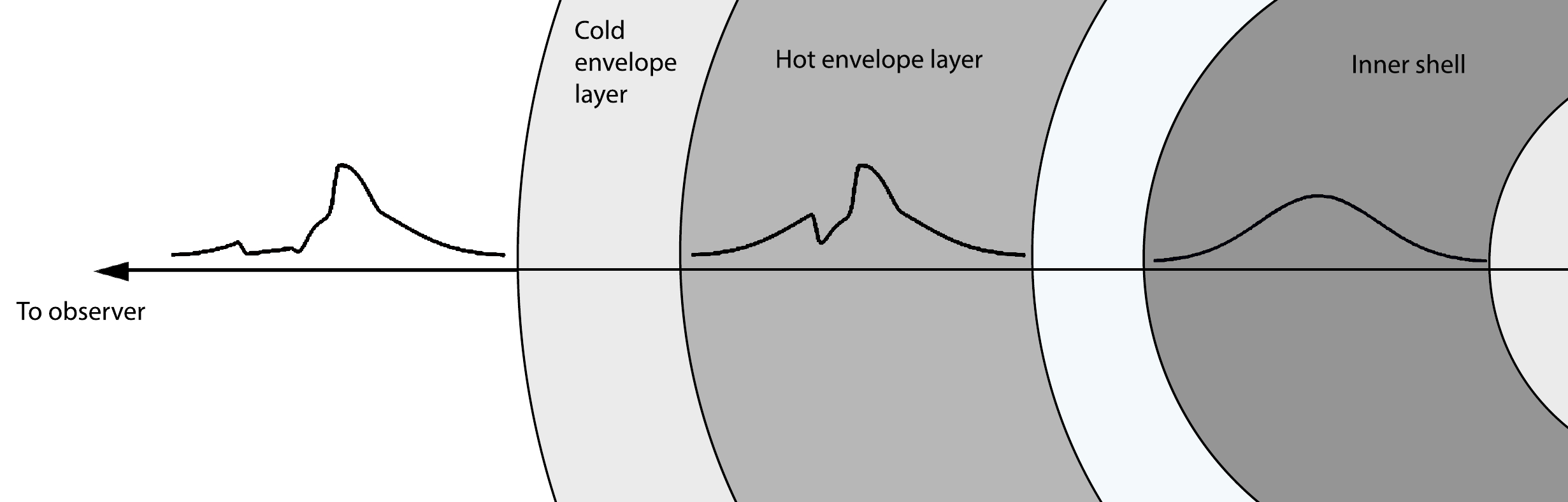}
\caption{The three H$\alpha$ emission lines plotted here represent the H$\alpha$ line at three different stages of its line of sight evolution.  In the background a not-to-scale cartoon representation of the physical structure that contributes to the given emission line.  The far right H$\alpha$ line is thermally broadened emission from only the inner shell.  The broad structure of the middle H$\alpha$ line is emission from only the inner shell, while the hot envelope contributes a P Cygni profile on top of the broad structure.  The far left H$\alpha$ line  is emission from the entire dsQN along the line of sight.  The radiatively cooled outer layer adds increased blue-side absorption to the H$\alpha$ line.}
\label{losHalpha}
\end{figure*}

In the case of SN 2006gy there exists numerous observations that detail its spectral evolution \citep{smith10}.  Unfortunately for other SLSNe, study of the spectra is limited by the amount of observations available, which in many cases is one spectral observation around the time of peak brightness.  As discussed in paper I, the dsQN can display an H$\alpha$ signature in which the underlying broad structure of the line is caused by thermally broadened emission originating from the inner shell, and emission and absorption from the envelope result in a blue-side absorption feature.  Fig. \ref{losHalpha} displays a schematic representation of the line of sight evolution of the H$\alpha$ line in a typical dsQN.   Unique to the dsQN scenario the wings of the H$\alpha$ line are caused by thermal broadening rather than velocity broadening (see right-most H$\alpha$ line in Fig. \ref{losHalpha}).  With this description the dsQN model can naturally account for the long-lasting breadth of the line that is inexplicable through other models \citep{smith10}.  As the H$\alpha$ line progresses through the hot component of the envelope a P Cygni profile is added to the structure of the line, this is displayed as the middle H$\alpha$ line in Fig. \ref{losHalpha}.  Finally the cold outer layer of the envelope adds absorption to the blue-side of the H$\alpha$ line (seen in the left-most line plotted in Fig. \ref{losHalpha}).

By classification SN 2005ap, PTF09cnd and PTF10cwr do not display any prominent hydrogen lines.  Although each of these SLSNe do show some indication of a weak H$\alpha$ line, as mentioned in the supplemental material associated with \cite{quimby11}.  Unfortunately the poor signal to noise (S/N) ratio eliminates any possibility of studying these weak lines.  The two spectral observations of SN 2008fz each show a strong H$\alpha$ line that resembles that of SN 2006gy \citep{drake10}.  While the H$\alpha$ lines display a broad underlying structure, the only spectral observation of SN 2008fz that is at an epoch which would show significant absorption by the envelope has the red-wing of the H$\alpha$ line cut off.  Without the red-side of the line there is no way of comparing the observations to our predicted dsQN H$\alpha$ line.  The spectra of SN 2008es contains broad Balmer lines which display a P Cygni profile however the poor S/N ratio of the observations inhibits any study of these lines \citep{gezari09}.

\begin{figure*}

\includegraphics[scale=1]{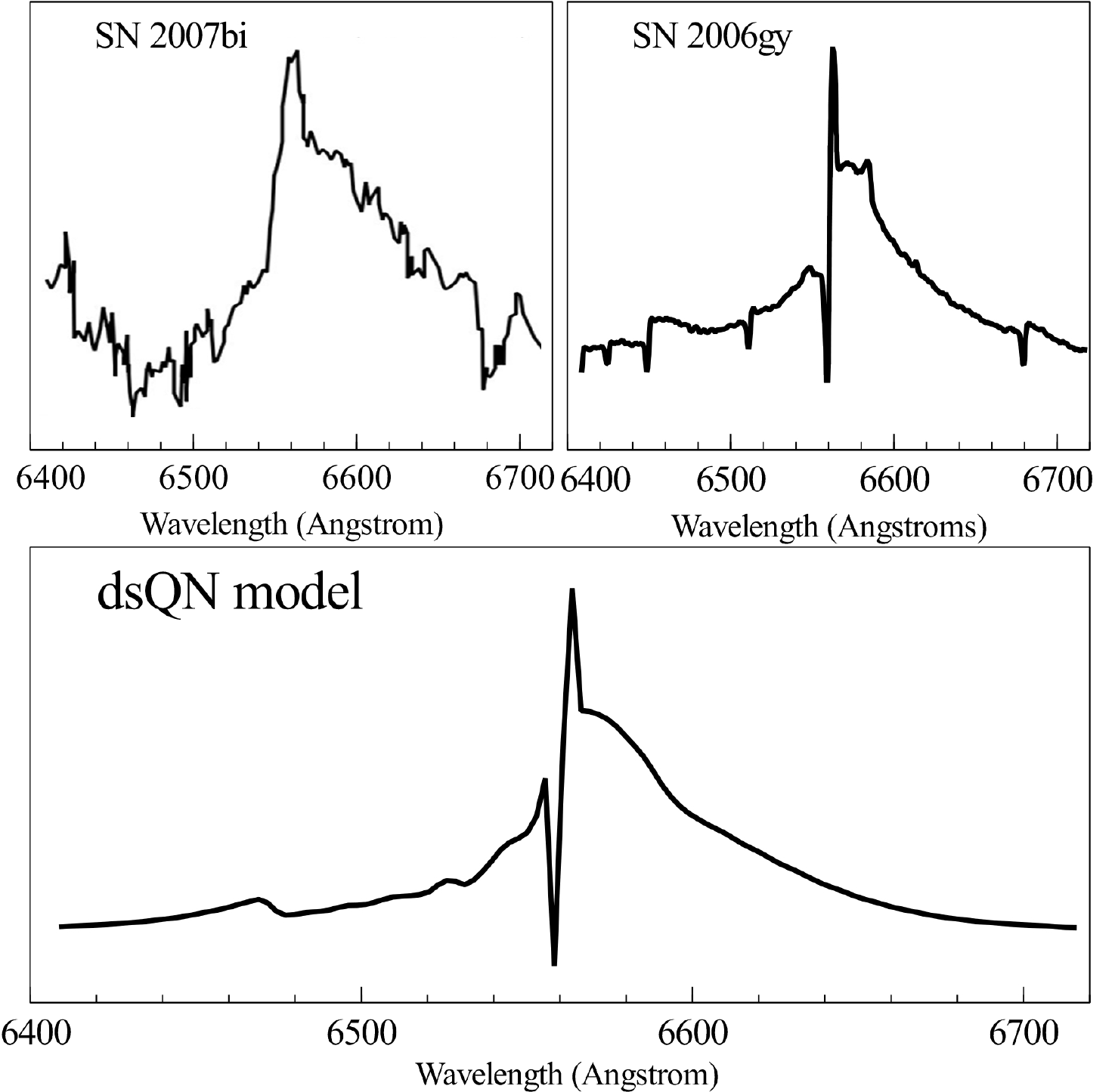}
\caption{Comparison of H$\alpha$ line from spectra of SN 2006gy, SN 2007bi and the dsQN model.  \textit{Top-Left:} Plotted is the observed H$\alpha$ line of SN 2007bi approximately 54 days after peak luminosity, data from Gal-Yam et al. 2009.  \textit{Top-Right:} The H$\alpha$ line from the spectrum of SN 2006gy observed at approximately 50 days post-peak luminosity, data from Smith et al. 2010. \textit{Bottom:} The dsQN model H$\alpha$ line 50 days post-peak luminosity, from paper I.}
\label{07biHalpha}
\end{figure*}

For our work in paper I we performed detailed analysis of the spectral evolution of the H$\alpha$ line observed in SN 2006gy.  The dsQN model was shown to provide a unique explanation for both the persistent broad structure of the line as well as the blue-side absorption feature seen in H$\alpha$ of SN 2006gy.  In this work we study the H$\alpha$ line of two other SLSNe; SN 2006tf and SN 2007bi.  For SN 2007bi the H$\alpha$ line observed in the spectra 54 days after peak brightness (see upper left panel of Fig. \ref{07biHalpha}) is mysterious as all other features of the spectra are consistent with a type Ic SN and thus should be free of hydrogen \cite{galyam09}.  For comparative purposes we have plotted the H$\alpha$ line observed in SN 2006gy (upper right panel of Fig. \ref{07biHalpha}) from a similar epoch as the spectral observation of SN 2007bi.  The bottom panel of Fig. \ref{07biHalpha} is the dsQN model H$\alpha$ line from the same epoch (from paper I).  Although the S/N ratio of the SN 2007bi prevents the study of the delicate structure of the H$\alpha$ line the large scale features are similar to that of a dsQN, namely the broad underlying structure and strong blue-side absorption.  

\begin{figure*}

\includegraphics[scale=0.75]{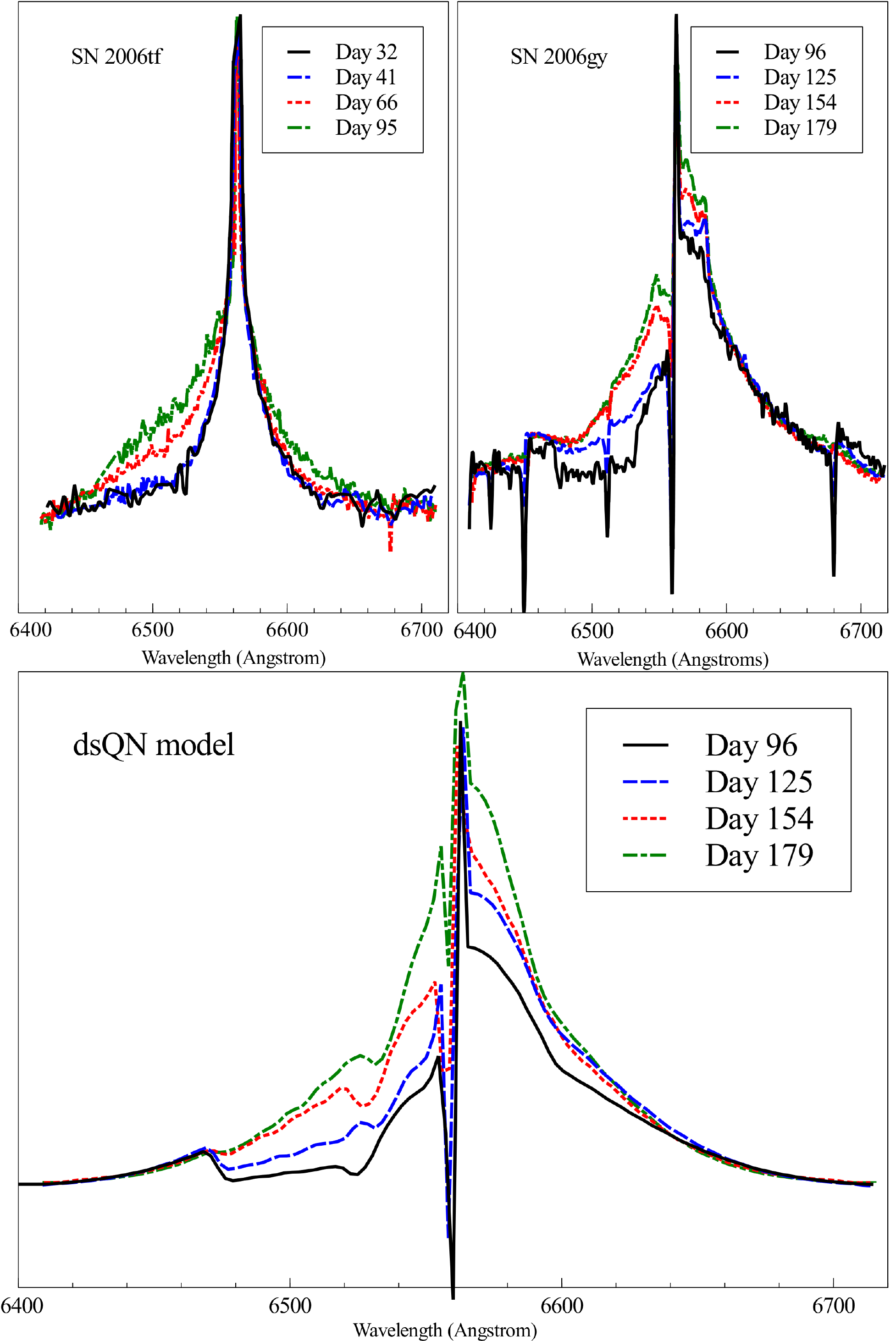}
\caption{Comparison of evolution of H$\alpha$ line from SN 2006tf spectra, SN 2006gy and our dsQN model H$\alpha$ line evolution.   \textit{Top-Left:} Plotted is an overlay of spectral observations of the H$\alpha$ line observed in SN 2006tf (data from Smith et al. 2008).  The observations are from; 32 (black solid line), 41 (blue dash line), 66 (red dotted line) and 95 (green dash-dot line) days after the first observation. \textit{Top-Right} An overlay of H$\alpha$ spectral lines from SN 2006gy is plotted, data from Smith et al. 2010.  The H$\alpha$ lines plotted are from days; 96 (black solid line), 125 (blue dashed line), 154 (red dotted line) and 179 (green dash-dot line) after the inferred SN explosion date.  \textit{Bottom:} Plotted is an overlay of the dsQN model H$\alpha$ line that was used to fit the observations of SN 2006gy in paper I.  The H$\alpha$ lines plotted are from days; 96 (black solid line), 125 (blue dashed line), 154 (red dotted line) and 179 (green dash-dot line) after the inferred SN explosion date.  }
\label{06tfHalpha}
\end{figure*}

The spectrum of SN 2006tf has multiple observations at different epochs along its light curve.  As noted by \cite{smith08} the H$\alpha$ line observed in SN 2006tf resembles that of SN 2006gy as it contains a broad underlying structure that becomes more prominent with time.  The upper left panel of Fig. \ref{06tfHalpha} displays an overplotting of four observations of the H$\alpha$ line of SN 2006tf.  For comparison the upper right panel of Fig. \ref{06tfHalpha} displays an overplotting of four spectral observations of SN 2006gy from similar epochs.  The bottom panel of Fig. \ref{06tfHalpha} displays dsQN model H$\alpha$ lines from roughly similar epochs as the observations of SN 2006tf and SN 2006gy.  In the dsQN model H$\alpha$ line and that of SN 2006tf and SN 2006gy there exists a broad component to the line as well as a blue-side absorption feature that diminishes in strength over time.  

The differences in the H$\alpha$ lines of SN 2006tf and SN 2007bi are consistent with the different $t_{\rm delay}$ of the dsQN description of these SLSNe.  In the case of SN 2007bi the shorter time delay ($t_{\rm delay}=18.5$ days) implies that the envelope is still dense at the time of spectral observation and thus there exists a strong blue-side absorption feature.  As for SN 2006tf the time delay is significantly longer ($t_{\rm delay}=22.5$ days) therefore the envelope is much more diffuse during the spectral observations and the blue-side absorption feature is weaker.

\subsection{The Envelope}
\label{envDisc}

The mass of the envelope as well as the initial shock temperature both act only to scale up or down the peak of the dsQN light curve.  We found that the only parameter capable of significantly changing the luminosity decay rate is the time delay between SN and QN.  All SLSNe with X-ray observations show remarkably quiet X-ray production.  This is expected in the context of a dsQN due to the cold outer layer of the envelope which would act as an efficient absorber of high energy radiation.  The formulation that we used for radiative transfer coefficients (see eqn. \ref{emi} and \ref{abs}) provided a good fit to each of the SLSN light curves that we studied.  This could be an indication that the continuum emission mechanism for these SLSNe resembles recombination which has a similar form as eqn. \ref{emi}.  Further study of SLSNe in which the light curve was observed in several passbands would help to determine the true temperature dependence for the emission coefficient and help in understanding the radiation mechanism.

\subsection{The Inner Shell}
\label{shellDisc}

Luminosity from the inner shell only begins to affect the shape of the light curves well past peak luminosity (typically once the luminosity has dropped to $\sim -19.5$ absolute magnitude).  The slowing of the luminosity decay rate, or \textit{plateauing} of the light curve in the late stage is caused by radiation from the inner shell shining through the diffuse envelope.

\section{Conclusions and Predictions}
\label{conc}
We have shown that the dsQN scenario can be used to explain the light curves of all eight SLSN targets studied.  In the context of dsQNe, progenitor stars ranging between 25-35 $M_{\sun}$ provide ample energy to power the large radiated energy budget of SLSNe. 

We found that the physical parameter with the greatest impact on dsQN light curve morphology was the time delay between SN and QN.  Shorter time delay dsQN yield a peak magnitude that is higher and a faster luminosity decay rate (narrower light curve).  While for longer time delays the peak magnitude is lower and the light curve is broader.  A variation in time delay in the dsQN description provides an explanation for the wide variety of SLSN light curve morphology.  From our analysis we found that for shorter time delay dsQN the inner shell may not be formed.  The implication being that the energy that would go into forming the inner shell may instead be lost to pressure-volume work, however further study of the dynamics of the QN-SN interaction must be undertaken.

We have also examined the singular H$\alpha$ spectral line profile found in three different SLSNe observations (SN 2006gy, SN 2006tf and SN 2007bi).  The broad structure of the line is accounted for by thermally broadened emission from the inner shell, while the intermediate peak and blue-side absorption feature are due to contribution from the envelope.  We found that the evolution of the blue-side absorption feature in the H$\alpha$ line of SN 2006gy and SN 2006tf is consistent with diffusion of the envelope.

Unique to the dsQN scenario is the fact that any core collapse SN that leaves behind a massive neutron star can in turn undergo a QN explosion.  This is due to the fact that the conditions of the interior of the progenitor star determine whether the neutron star could become susceptible to QN collapse.  There is no correlation to the progenitor star envelope (for example whether or not hydrogen is present), thus we expect a wide variety of types of SN can become super-luminous due to re-brightening via a QN collision.

A distinguishing feature predicted by the dsQN model is a unique chemical abundance caused by the spallation of the SN envelope by the QN ejecta \citep{ouyed11}.  Recent observations have suggested plausible signs of the QN were found in Cas A \citep{hwang12}.  As described in \cite{ouyed11} the layer of the SN envelope undergoes spallation by the QN ejecta depends on the density of the envelope at the moment of impact.  For shorter time delay QN the Ni layer would be destroyed in favour of the production of sub-$^{56}$Ni elements such as; Ti, V, Cr and Mn \citep{ouyed11}.  This process should also lead to mixing that would cause Fe to be found in the outer regions of a dsQN remnant \citep{ouyed11}.  The formation of $^{44}$Ti at the expense of $^{56}$Ni will play an important role in the very late stage ($t\sim 1000$ days) luminosity of dsQN.  For longer time delay QN the spallation process occurs in the carbon \& oxygen layer of the SN envelope which leads to a unique chemical signature of the dsQN.  Spallation of the carbon \& oxygen layer would lead to an over-abundance of lithium in the dsQN remnant \citep{ouyedCEMP}.

A dsQN is expected to emit two bursts of X-rays.  The first X-ray emission event would occur when the shock from the original SN breaks out of the stellar envelope and the second analogously occurs for the QN shock break-out of the SN envelope.  If the time delay between SN and QN is short then the X-ray bursts could in fact be overlapped leading to a broadened X-ray light curve.  However if the time delay is long then there should be two distinct X-ray peaks.

\footnotesize{
\bibliographystyle{mn2e}
\bibliography{multibib}
}

\label{lastpage}
\end{document}